\documentclass[sigconf]{acmart}
\usepackage{xspace}
\usepackage[ruled,vlined]{algorithm2e}
\usepackage{amsmath}
\hypersetup{draft}

\DeclareMathOperator*{\argmax}{arg\,max}

\newcommand{\name}{SINVAD\xspace}

\AtBeginDocument{%
  \providecommand\BibTeX{{%
    \normalfont B\kern-0.5em{\scshape i\kern-0.25em b}\kern-0.8em\TeX}}}

\copyrightyear{2020}
\acmYear{2020}
\setcopyright{acmcopyright}\acmConference[ICSEW'20]{IEEE/ACM 42nd International Conference on Software Engineering Workshops }{May 23--29, 2020}{Seoul, Republic of Korea}
\acmBooktitle{IEEE/ACM 42nd International Conference on Software Engineering Workshops (ICSEW'20), May 23--29, 2020, Seoul, Republic of Korea}
\acmPrice{15.00}
\acmDOI{10.1145/3387940.3391456}
\acmISBN{978-1-4503-7963-2/20/05}
\begin{document}

\title{\name: Search-based Image Space Navigation for DNN Image Classifier Test Input Generation}

\author{Sungmin Kang}
\affiliation{%
  \institution{School of Computing, KAIST}
  \city{Daejon}
  \country{Republic of Korea}
}
\email{stuatlittle@kaist.ac.kr}

\author{Robert Feldt}
\affiliation{%
  \institution{Chalmers University}
  \city{Gothenburg}
  \country{Sweden}
}
\email{robert.feldt@chalmers.se}

\author{Shin Yoo}
\affiliation{%
 \institution{School of Computing, KAIST}
 \city{Daejon}
 \country{Republic of Korea}
}
\email{shin.yoo@kaist.ac.kr}

\newcommand{\todoc}[2]{{\textcolor{#1} {\textbf{[[#2]]}}}}
\newcommand{\todored}[1]{\todoc{red}{\textbf{[[#1]]}}}
\newcommand{\sm}[1]{\todoc{blue}{sungmin: #1}}
\newcommand{\sy}[1]{\todoc{red}{SY: #1}}
\newcommand{\rf}[1]{\todoc{orange}{RF: #1}}

\begin{abstract}
The testing of Deep Neural Networks (DNNs) has become increasingly 
important as DNNs are widely adopted by safety critical systems. While many 
test adequacy criteria have been suggested, automated test input generation 
for many types of DNNs remains a challenge because the raw input space is too 
large to randomly sample or to navigate and search for plausible inputs. 
Consequently, current testing techniques for DNNs depend on small local 
perturbations to existing inputs, based on the metamorphic testing principle.
We propose new ways to search not over the entire image space,
but rather over a plausible input space that resembles the true training distribution.
This space is constructed using Variational Autoencoders (VAEs), and navigated through
their latent vector space. 
We show that this space helps efficiently produce test inputs 
that can reveal information about the robustness of DNNs 
when dealing with realistic tests, opening the field to meaningful exploration through the space of highly
structured images.

\end{abstract}


\maketitle

\section{Introduction}
\label{sec:intro}

The use of convolutional neural networks for image classification
\cite{Krizhevsky2012alexnet} has sparked significant growth in the use of such
techniques; for example, Deep Neural Networks (DNNs) are widely used in 
computer vision, natural language processing, and playing games~\cite{LeCun2015DL}. 
As DNNs can greatly enhance the accuracy of computerized perception, their use 
even in safety-critical domains, such as autonomous driving~\cite{Chen2017aa,chen2015deepdriving} 
or medical imaging~\cite{Litjens2017rt}, has also seen a rise. 

However, despite the increasing adoption of DNNs in these domains, 
many of their safety properties are difficult to 
comprehend or fix. For example, adversarial examples~\cite{CarliniW16a,PapernotMJFCS15,Kura1607,Goodfellow43405} add a
highly targeted noise vector to normal inputs, causing DNNs to
deviate from human perception. Such attacks are not only also possible by 
manipulating real world physical objects that intereact with DNNs~\cite{Kura1607}, 
but also possible for other domains such as speech recognition~\cite{Schonherr2018if} 
and machine comprehension of texts~\cite{Jia2017hs}.

As the necessity of thorough testing of neural networks has received wide
acknowledgement, various methods have been proposed to fix this, particularly
through the traditional approach of coverage. Pei et al. \cite{Pei2017qy} track the
activation status of each neuron, and argues that tests should aim to activate
a diverse set of neurons. Ma et al. \cite{Ma2018aa} posit that test suites
should have each neuron emit multiple
strata of outputs. Kim et al. \cite{Kim2019aa} quantify how surprising an unseen 
input is and argues that effective test suites should use inputs with differing levels of surprise.

While the number of approaches to create adversarial and difficult examples has grown, 
there has been relatively little effort in
attempting to search the larger space of valid image representations. 
It is important to search this space, as images made by tweaking a few pixels
without regard to what images are realistic are often not helpful. 
For example, even if we find that perturbed pixel values within a certain image region cause the neural
network to misclassify as presented in DeepXplore~\cite{Pei2017qy}, it is unclear
what one can do to fix this and whether it matters, as such a perturbation is unlikely to
appear in practice. Ideally, we would like to identify real-world scenarios in which 
the classifier fails; this is possible when one searches over the space of valid
image representations, not when search is performed over small perturbations of an
existing image.

In this work, we present \name (Search-based Input space Navigation using Variational AutoencoDers). \name navigates in the space of plausible images, a subspace
of images that semantically resemble those of a specific dataset, by
leveraging Variational Autoencoders (VAEs)~\cite{Kingma2014VAE}, 
a class of generative models. 
It is first shown that, as argued beforehand, optimization over the entire
image space yields static-like noise images that are challenging
to interpret, while optimization over the latent space of a 
VAE yields realistic images that may appear as a harmless input, and
thus are related with the dataset in question. Equipped with this validation of our initial
premise, we subsequently perform a guided 
search for plausible images that reside close to the decision boundary, 
enhancing our ability to perform boundary testing using images that lie on the
semantic boundary of labels, found though \name. 

We further verify that search is not only possible on one DNN, but can be adapted
so that multiple DNNs provide guidance to search. Specifically, we employ a 
GA-based technique to uncover the differences between neural networks within 
the space of plausible images. 
By analyzing the characteristics of images that cause judgement differences 
between neural networks, one may uncover plausible faults and
differences in neural networks. 

Finally, leveraging certain semantic properties (e.g. it is difficult to draw a 3 that looks like a 4), and that our test data generation
algorithm closely emulates such characteristics, a test case generation technique
that picks up pairs of categories which are at risk of being confused with each other is provided. 
We show that the results
of the tool roughly correlate with the semantic separation between categories, 
and further verify that when training data labels are mixed, our technique
picks up such change faster than the default test data can.

The contribution of the paper is as follows:
\begin{itemize}
  \item We introduce \name, a novel method to search through the 
space of plausible images using a neural network based representation of images,
which is shown to be a valid way of searching for images that meet desiderata while
remaining plausible;
  \item We evaluate border images, images optimized to be confusing to the neural network, to show that dropout-induced models 
indeed provide widely different predictions between each other, to a greater 
extent than standard dataset test images; a qualitative assessment of generated
border images confirms that they are semantically `in between';
  \item We present a technique using the test generation algorithm that produces
tests capable of identifying training problems earlier than the provided test set. 
\end{itemize}

The remainder of this paper is organized as follows. 
Section~\ref{sec:background} introduces the related literature, providing 
background to this paper. Section~\ref{sec:approach} presents nomenclature, and defines the problem of searching the space of DNN inputs in a formal way.
How \name performs search in the large DNN input space is subsequently explained.
Section~\ref{sec:rqs} presents the research questions we aim 
to answer with experiments that are described in 
Section~\ref{sec:experiments}. 
Section~\ref{sec:threats} presents threats to validity, and 
Section~\ref{sec:conclusion} provides discussion and concludes.

\section{Background}
\label{sec:background}

This section presents background information relevant to \name. 

\subsection{Testing of DNNs}

As the need to test and validate DNN models increases, many recent work focused on testing techniques for DNNs. The early work mostly concerns how to differentiate good and bad inputs, i.e., how to choose inputs that are more likely to reveal suboptimal behavior of a given DNN. Neuron Coverage (NC) was the first test adequacy metric proposed for DNNs: it computes the ratio of neurons in a given DNN that were activated above a predetermined threshold value by a set of test inputs~\cite{Pei2017qy}. Many different adequacy criteria based on neuron activation have since been proposed, including k Multisection Neuron Coverage (kMNC), a finer granularity refinement of NC, and Strong Neuron Activation Coverage, a criterion that focuses on out-of-distribution activations (i.e., ones that are beyond the range of activations observed during training)~\cite{Ma2018aa}. Surprise Adequacy (SA) subsequently focused on the similarity between neuron activations observed during training (aggregated as probability distributions) and a single input, allowing the tester to measure how \emph{surprising} a new input is, and consequently how likely it is to reveal unexpected behavior~\cite{Kim2019aa}. We use SA to guide our search for new inputs for DNN based image classifiers in this paper. Zhang et al.~\cite{zhang2018DeepRoad} use GANs to generate realistic road images for autonomous driving applications. 

In addition to dynamic testing, there are work that try to verify DNNs with a correctness guarantee. ReluPlex extends the simplex method, applying SMT solvers to provide safety guarantees for DNNs~\cite{Katz2017ad}. Huang et al. proposed a verification technique for image recognition DNNs, checking whether images within a certain distance of each other are classified identically~\cite{Huang2017kx}. A general discussion of testing machine learning techniques is provided in Zhang et al.~\cite{zhang2019machine}.

\subsection{Difficulties of Input Search for DNNs}

One of the biggest challenges in searching for new test input for DNNs is the semantic manifold problem~\cite{Yoo2019aa}. Only a miniscule fraction of the entire input space of a image classifier constitutes the space of valid inputs that are meaningful images; the vast majority of the space is uninterpretable noise. However, it is not possible to clearly define which regions of the overall input manifold are semantically meaningful to humans, preventing us from navigating the space freely guided by fitness functions. Most of the existing testing techniques for DNNs circumvent this issue by adopting the metamorphic testing principle: given a seed image, a DNN under test should behave the same after we either inject a small amount of noise (e.g., adding noise that humans can safely ignore), or apply semantically irrelevant changes (e.g., synthesizing an image with the same semantic content but under a different weather condition) ~\cite{Pei2017qy,Tian2018aa,Kim2019aa}. \name is an attempt to search in the space of semantically meaningful inputs, by using VAEs as an approximation of the space of meaningful inputs.

\subsection{Generative Models}
As DNNs became more powerful function approximators, generative
models that can imitate complex distributions became more plausible to implement. The core
theme of generative models is that if one can model a distribution to calculate
the odds of a certain input, one can also sample from that distribution to
make plausible inputs. The advent of machine learning has brought many powerful
generative models that approximate distributions, either explicitly or
implicitly. Generative Adversarial Networks~\cite{Goodfellow2014GAN} employ a generator
and discriminator network, in which the generator attempts to match the input
distribution while the discriminator tries to find the discrepancies between
the generator-approximated distribution and the true distribution. The training
process implicitly guides the generator network to resemble the true
distribution. PixelCNNs~\cite{Oord2016PixelRNN} model the dependencies between
neighboring pixels, sampling each pixel conditionally on nearby pixels. 
PixelCNNs are trained explicitly to assign
high probability to provided images. Variational AutoEncoders (VAEs)~\cite{Kingma2014VAE} do not explicitly
calculate the odds of an image, but maximize the evidence lower bound (ELBO),
which acts as a lower bound of the odds of an image. There are many
other generative models; one may look to Foster~\cite{Foster2019gdlbook} for further examples. For our
purpose, any generative model with latent variables (e.g. GANs or VAEs) 
that can function as a condensed search space may be used.

\begin{figure*}[h!]
  \centering
  \includegraphics[width=\textwidth]{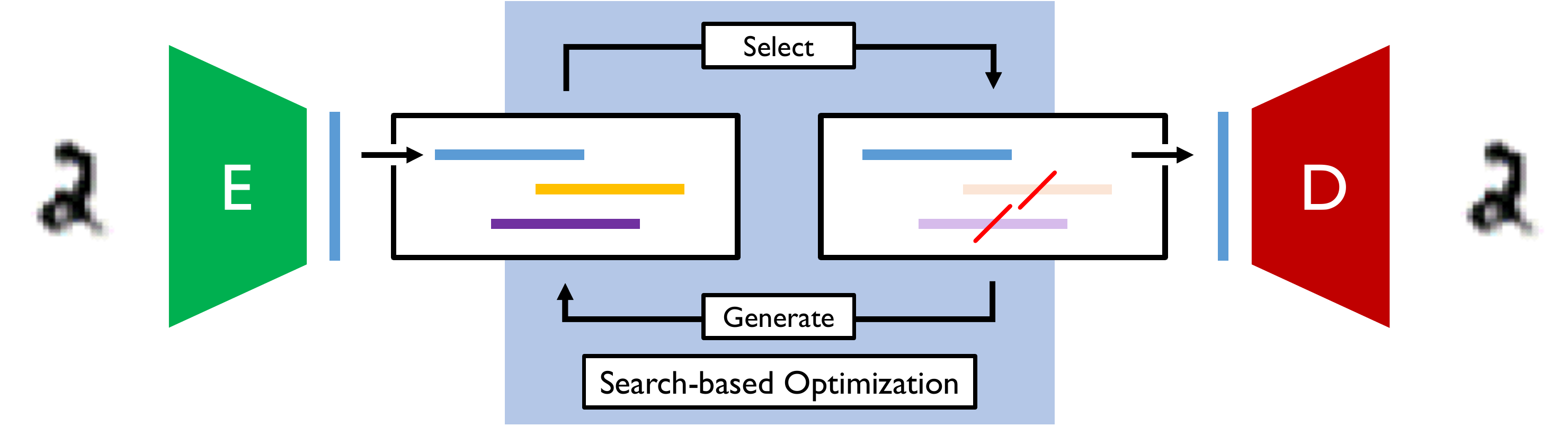}
  \caption{Diagram of \name. \name uses the VAE encoder (E) to encode
  images into a \textit{latent space} (bars). Search-based optimization is performed
  not on raw pixels, but on latent representations. A representation is
  transformed back to an image using the VAE decoder (D). The use of VAE 
  networks before and after keeps search within the subset of plausible images.}
  \label{fig:diag}
\end{figure*}

\section{\name: Search-based Input Space Navigation using VAEs}
\label{sec:approach}

\def \imgI {\mathcal{I}}
\def \catP {\mathcal{P}}
\def \dataD {\mathcal{D}}

\subsection{Problem Statement}
\label{sec:problem}
A common application of DNNs is as image classifiers.
In such applications, a neural network $N$ acts as a function that maps the space of images
$\imgI = [0, 1]^{c \times w \times h}$ to a space representing the probability of each class
$\catP = \{v \in [0, 1]^n : \sum_{i=1}^n v_i = 1\}$, where there are $n$ categories. 
Formally, the DNN is a mapping $N_p: \imgI \rightarrow \catP$, and a mapping
$N_c(i) = $one-hot$( \argmax N_p(i))$ is derived.

In practice, the semantic meaning of most images is unclear, so
the neural network is trained on a subset of images from the space of plausible
images $\dataD$ and maps to $\catP$. 
In test generation, this becomes a problem as the space of all images of a certain resolution 
is usually much larger than the space of all such images that are valid $\dataD$. If one looks at the MNIST
dataset for example, it is clear that the space of grayscale 28 by 28 images is
much larger than the space of digit images, for most random grayscale
images look more like static noise than coherent numbers~\cite{Yoo2019aa}. 
Then, due to the vast
size difference between $\imgI$ and $\dataD$ and the complexity of the neural
networks employed to solve classification problems, it is often easy to obtain
images that meet a certain criteria that are not particularly interesting or meaningful. 
For example, it is relatively easy to construct an image that resembles noise, but
is classified with high confidence as a cheetah~\cite{Nguyen2015aa}. 
While this is
interesting, our concern is often more closely related to the implications of
when neural networks are released in the wild, e.g. in what situations would
there be a risk of failure? As such, it would be beneficial if we could search
exclusively in the space of plausible inputs.

\subsection{\name}
We propose to use DNN based generative models,
such as VAEs or GANs, to solve this problem. 
A schema of our approach \name is provided in Figure \ref{fig:diag}.
Generative models estimate the distribution of a provided dataset over the entire
space of all images $\imgI$. In particular, VAEs effectively try to find a mapping
between $\mathbb{R}^d$ and $\dataD$; the encoder ($E: \dataD \rightarrow \mathbb{R}^d$)
maps from images to latent representations, and decoders ($D: \mathbb{R}^d \rightarrow \dataD$)
operate vice versa. 
By executing traditional search-based optimization algorithms,
but within the latent representation space, we can restrict our search space to
$\dataD$, instead of the full image space $\imgI$. Hence the result of search 
will mostly be images that closely resemble true images from $\dataD$.
Note that since the VAE itself has only been trained on a small sub-sample of $\dataD$, i.e. the training set, it might not be a very good mapping to\slash from the actual $\dataD$. We thus have to investigate the practical value of our approach empirically, in the following.

\newcommand{\rqonename}{Plausibility\xspace}
\newcommand{\rqtwoname}{Indecisiveness\xspace}
\newcommand{\rqthreename}{Differential Testing\xspace}
\newcommand{\rqfourname}{Application\xspace}

\section{Research Questions}
\label{sec:rqs}

In this paper, we seek to find the answer to four research questions.

\paragraph{\textbf{RQ1. \rqonename:} Can \name generate images closely 
related to $\dataD$?} Our justification for using complex generative models
is the assumption that, if employed to reduce the search space, the resulting images will be more closely aligned with the
true data distribution $\dataD$, i.e. more `realistic' than similar techniques
that do not employ generative models. To verify this point, we compare
\name with search-based optimization algorithms that do not use VAEs. We qualitatively compare
whether an image is realistic or not, as human perception is the only true
guide in this case. 

\paragraph{\textbf{RQ2. \rqtwoname:} Can \name generate images near the decision boundaries of 
neural networks?} 
Using \name with a certain fitness function, 
one can make VAE-generated images that are near oracle decision
boundaries, i.e. the category identity of generated images is unclear even
to humans. We verify whether generated images are also close to neural network
decision boundaries by employing dropout. When certain neurons are dropped,
neural networks are more likely to make mistakes on confusing inputs rather
than straightforward inputs. Using this, we aim to observe that our images are indeed
more difficult to classify than default test sets, thus providing additional
scrutiny while testing. 

\paragraph{\textbf{RQ3. \rqthreename:} Can \name perform differential testing for neural network
comparison?} 
To further investigate the potential of the image space search that \name enables, we use it for differential testing of neural networks. By traversing the space and comparing the outputs of the networks, we investigate whether we can identify meaningful semantic differences in their behavior. In turn, this may highlight important properties of DNNs under test.
\paragraph{\textbf{RQ4. \rqfourname:} Can \name be utilized to identify 
`weak spots' of neural network classifiers?} 
Using \name, we attempt to find pairs of categories that 
a given DNN based image classifier frequently confuses. 
For example, we would like to answer whether a given image classifier is more 
easily confused between the pair of 0s and 1s, or 4s and 9s. Such information can reveal
potentially harmful pairs, indicating where the classifier needs more
data and further training. If we ask \name to find an image that looks like
class X but is classified as Y, the success rate of this optimization differs by pair. 
We posit that the success rate is correlated with the risk of the 
DNN under test making a mistake, and verify this by artificially mixing
labels and checking whether the genetic algorithm's success rate increases
as the labels become more mixed. 

\section{Empirical Analysis}
\label{sec:experiments}

\subsection{Experimental Setup}
\label{sec:setup}

Two datasets are used throught the paper. MNIST~\cite{LeCun2010vg} is a grayscale digit image
dataset, each $28 \times 28$ pixels large. There are 50000 training, 10000 validation, 
and 10000 test images, but in this paper all training and validation images are
used for training. SVHN~\cite{Netzer2011SVHN} is a dataset containing real-world digits
obtained from Google Street View images. There are 73257 training images and 26032
tests. Each image is $32 \times 32$ pixels large, with RGB colorization. 

For MNIST, a VAE that has one hidden layer of size 1600 for both encoder and decoder 
structures is used. The encoding vector size is 400. %
For SVHN, we employ a network with 8 hidden layers
in the encoder and decoder respectively. The larger VAE for SVHN is needed
due to the greater difficulty of modelling images with background variation
and color schematics. %
VAEs for both datasets are trained for 50 epochs, which was sufficient for training
to convergence. 

Regarding the neural networks to emulate the DNNs under test, a convolutional neural 
network with 6 hidden layers is used for MNIST. 
A network of an overall similar
architecture with 9 hidden layers is used for SVHN. In Section \ref{sec:expr3}, we use
a VGG-19~\cite{Simonyan15VGG} network to perform differential testing with our 9-layer model.
The likelihood-based surprise adequacy (LSA) metric is calculated using the activation traces 
of the penultimate layer, i.e. the input vector to the final neural network layer which produces
softmax logits. %

\subsection{\rqonename (RQ1)}
\label{sec:expr1}
Through this experiment, we seek to answer RQ1: whether VAE-based optimizations
yield more `realistic' result images. To verify this point, we perform the 
following experiment. We attempt to obtain images that have a target Surprise
Adequacy~\cite{Kim2019aa}. To do this, 
we perform a variant of random hill climbing in which we
add a random value sampled from the standard normal distribution to a single
element in the vector representation and compute fitness
based on the function $f(i) = |LSA_{N}(i)-t|$, where $LSA_N$ is LSA calculated
using a neural network $N$, and $t$ is a target SA value. 
If the fitness is lower, we keep
the change and move on to the next element, modifying each element one by one in order. 
On raw pixels, we modify a precursor representation $r$ which has the same dimension
as normal images. The image from representation is obtained by applying
the $tanh$ function elementwise on $r$. This normalization restrains the values
of pixels to be within a certain range, so that the image is within the typical image
space. 
For the VAE representation, we modify a 
latent vector $z$ and obtain the image through $D(z)$. 
Search starts from a random representation so that comparison is fair. Specifically, $r$
is generated by sampling from unif$(0, 1)$ for each element, while $z$ is generated
by sampling from the standard normal distribution $N(0, 1)$ for each element.

\begin{figure}[h!]
  \centering
  \includegraphics[width=\columnwidth]{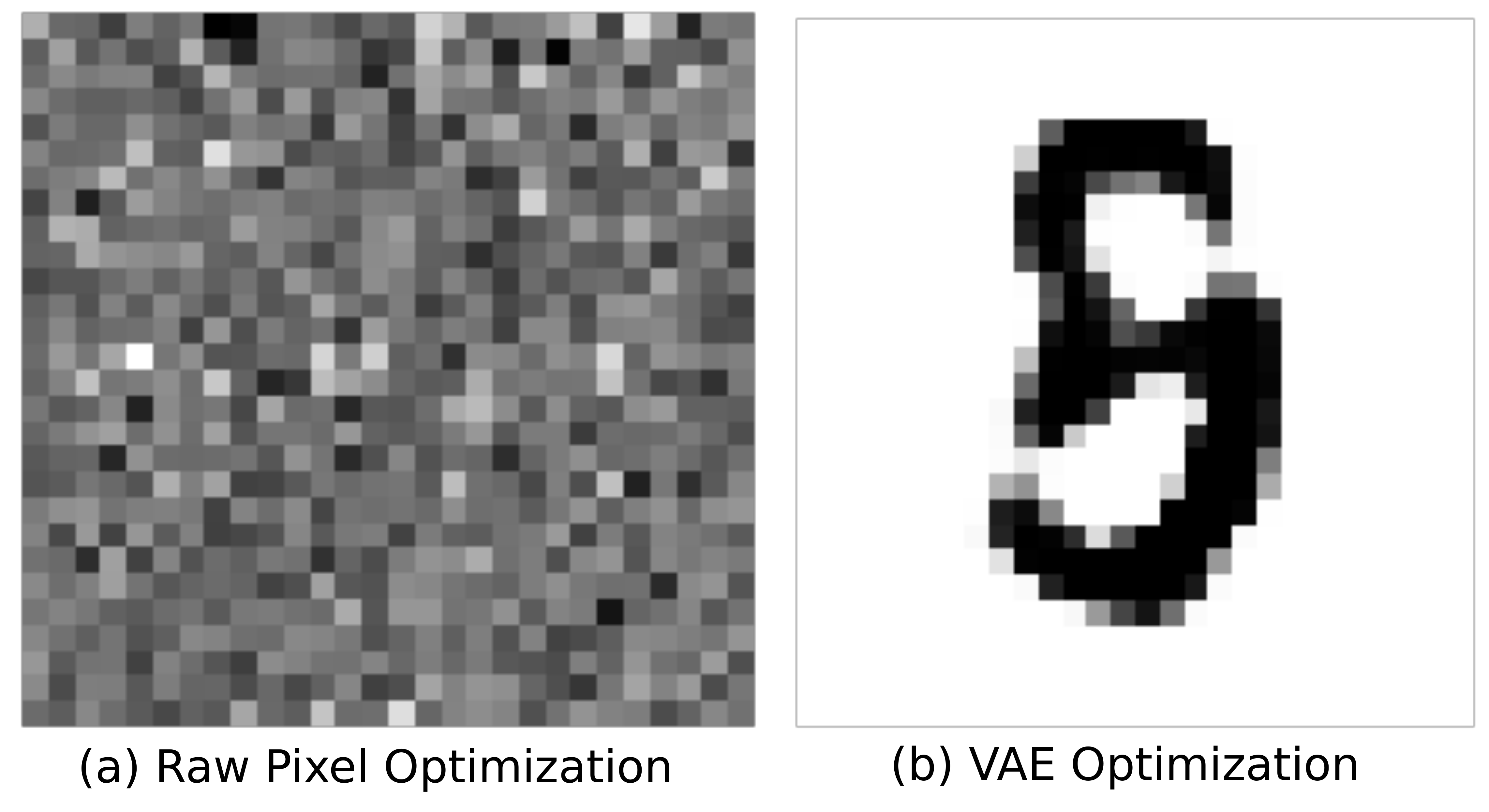}
  \caption{Images produced by optimization either at the raw, pixel level (a, left) or using \name (b, right). Despite having the same fitness value the VAE based image looks closer to a ``real'' digit.}
  \label{fig:rawVvae}
\end{figure}

A result is presented in Figure \ref{fig:rawVvae}; while both images achieve
similar performance in terms of the fitness function, the raw pixel-based 
optimization result appears to lack any global structure, whereas the
optimization guided by a VAE yields images with coherent structure that indeed
appear like a digit.

\begin{figure}[h!]
  \centering
  \includegraphics[width=\columnwidth]{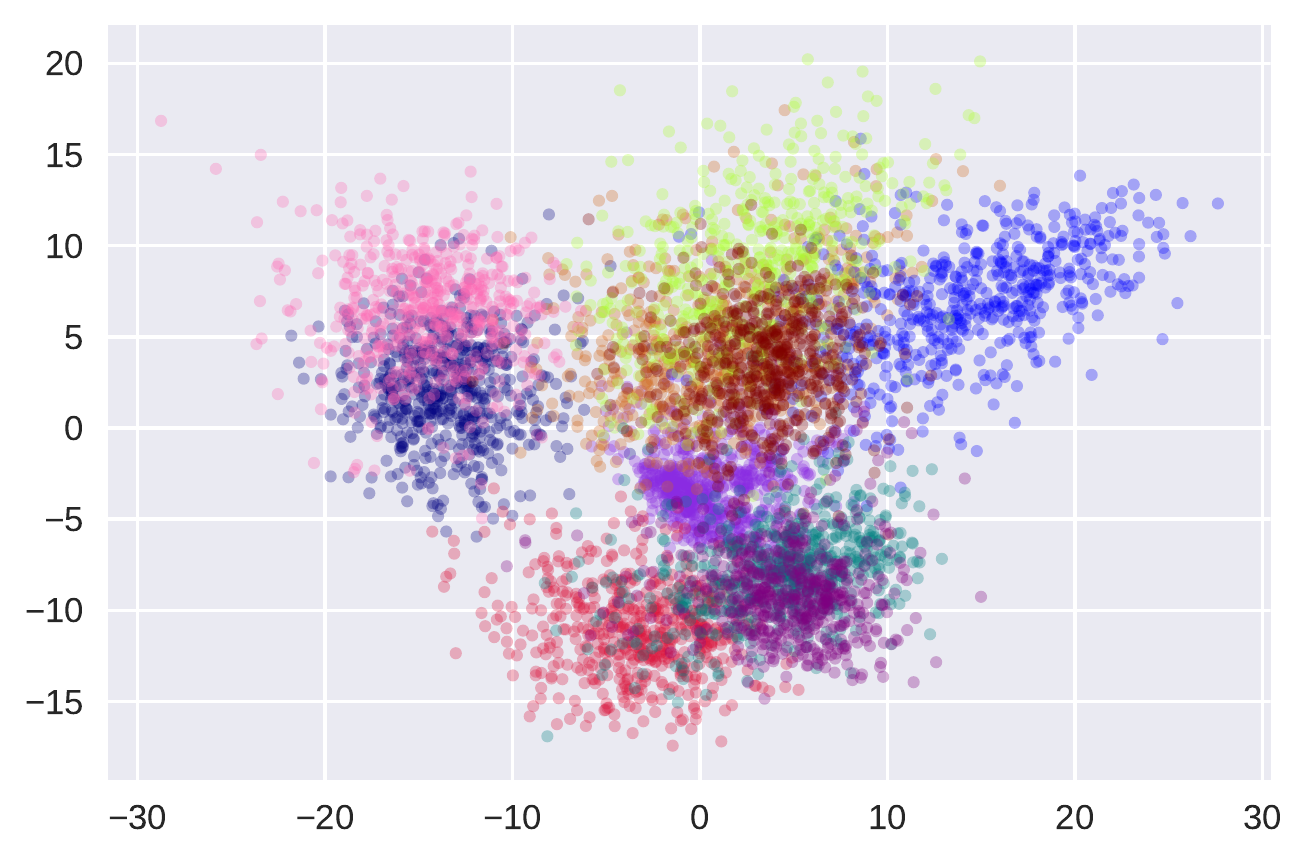}
  \caption{PCA projection of training data AT space, 
  with each color signifying ATs from different digits.}
  \label{fig:pca}
\end{figure}

We may also look at these results through the prism of activation trace (AT)
space~\cite{Kim2019aa}. %
The activation trace of an image is the set of neuron activations that are
elicited when it is provided as input. While neural networks are 
high-dimensional and thus it is difficult to directly observe AT space,
we may perform dimension reduction techniques such as PCA to get a glimpse
of the structure. In turn, we may observe where each image is positioned
in the AT space, providing a visual persepective on `familiarity' to the
neural network. This paper uses PCA as it allows projection of new data. 
Figure \ref{fig:pca} depicts the PCA projection of the AT of
training data, each color representing the AT of a certain digit. 

\begin{figure}[h!]
  \centering
  \includegraphics[width=\columnwidth]{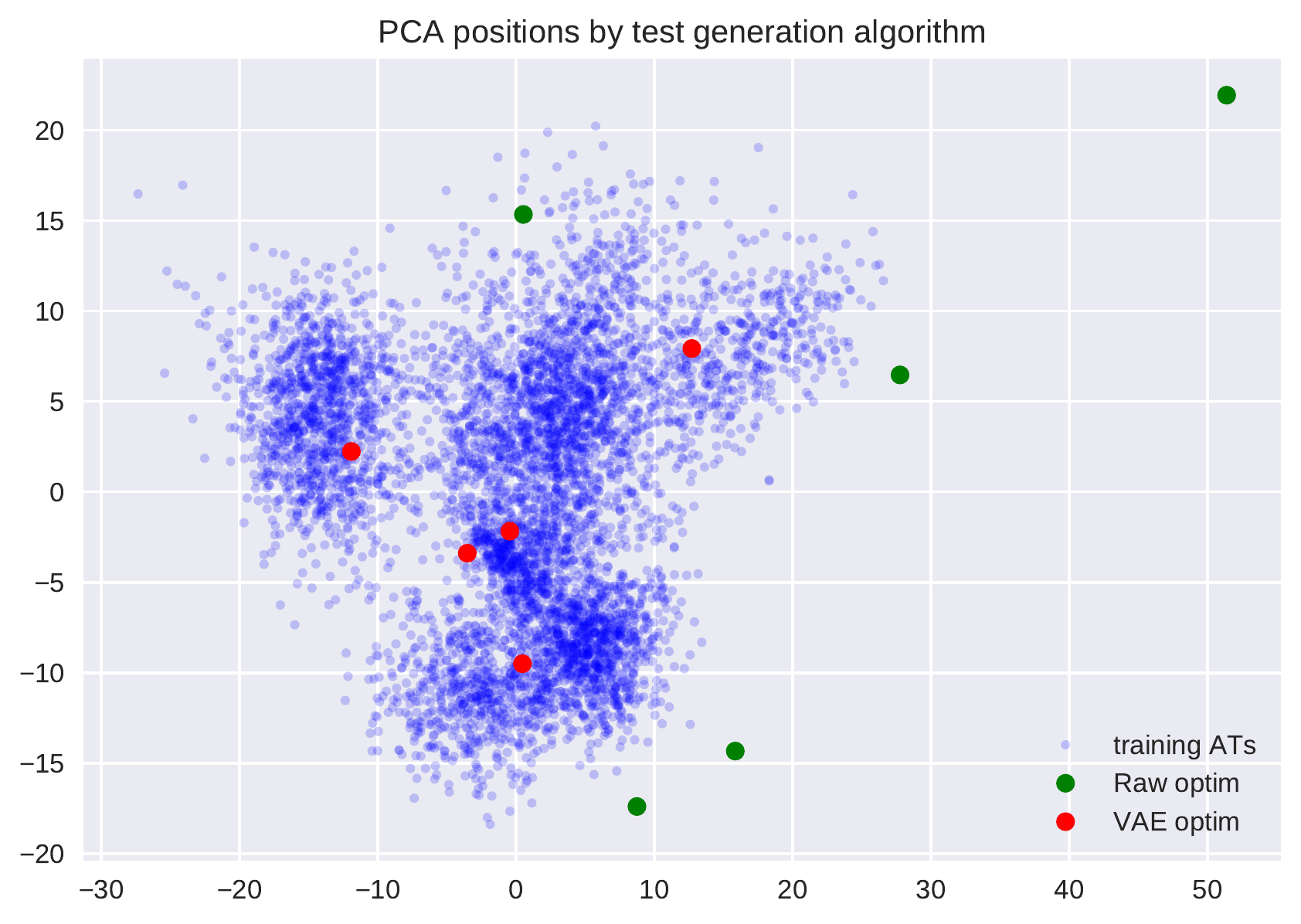}
  \caption{Placement of images optimized either at the raw, pixel level (green) or at VAE latent vector level (red), in relation to the PCA projection of the training images.}
  \label{fig:pca_adv}
\end{figure}

Using this projection, we can visualize where both VAE-based images and 
raw pixel search based images are, to check where they are positioned in the 
space. Figure \ref{fig:pca_adv} shows an example of such a visualization. As one can 
see, the GA-based images are closer to the true data distribution,
while raw pixel based images end up around the outskirts of the real image
activation trace distribution. %

\begin{figure}[h!]
  \centering
  \includegraphics[width=\columnwidth]{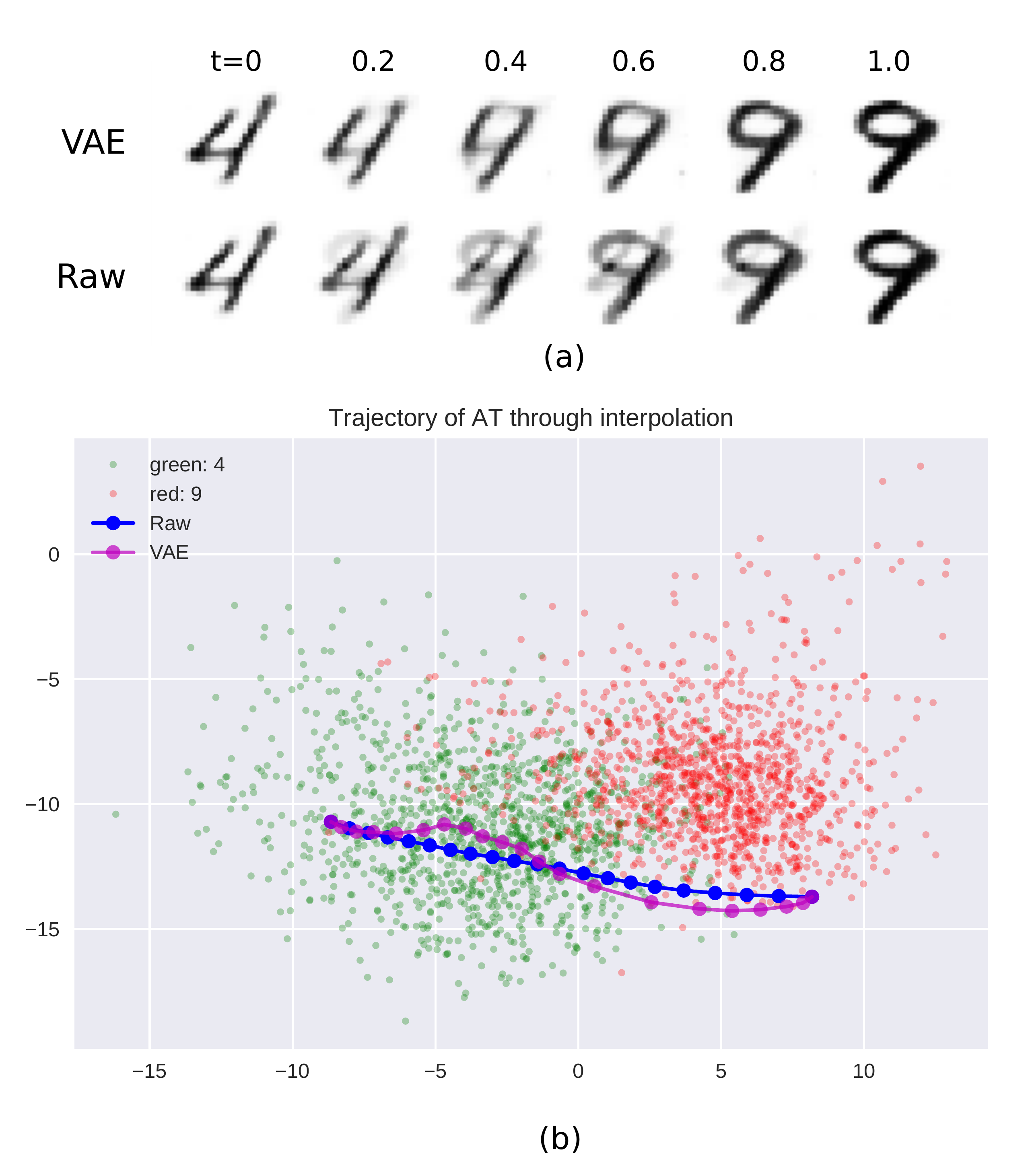}
  \caption{Trajectories of interpolated images projected into the PCA space.}
  \label{fig:interpolation}
\end{figure}

Using the visualization properties of the PCA projection, we can further illustrate
how search can be performed over the VAE latent space, and how this can yield more plausible
results than searching over raw pixels. Given two images, we
interpolate over both raw pixels and the latent space; specifically, we
obtain raw pixel interpolations for a certain pixel $[i, j]$ through
$i_t[i, j] = (1-t)*i_0[i, j] + t*i_1[i, j]$, while 
VAE interpolated images for image $i_0$ and $i_1$ are obtained with:
$i_t = D((1-t)*E(i_0)+t*E(i_1))$. We compare these interpolated images
side by side in Figure \ref{fig:interpolation}(a), while their PCA trajectory in the AT space
is presented in Figure \ref{fig:interpolation}(b). Note that in Figure \ref{fig:interpolation}
(a), the raw representation has no understanding of semantics, hence it mixes
the images unnaturally. On the other hand, the VAE representation tries to
naturally interpolate between the images, keeping intermediate images somewhat
plausible. This effect also appears in Figure \ref{fig:interpolation} (b); 
while the raw representation just takes a straight trajectory from the first
to last point without regard of semantics, the VAE representation spends little
time in the implausible regions where there are few previous images and more time
where there is a dense distribution of real images. This again testifies that
VAEs can generate plausible images more consistently. 

\begin{figure}[h!]
  \centering
  \includegraphics[scale=0.2]{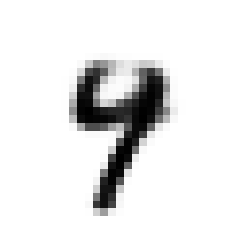}
  \caption{An ambiguous image generated through GA optimization using 
  Eq. \ref{eq:bound_fitfunc} as a fitness function.}
  \label{fig:ambiguous}
\end{figure}

By changing the fitness functions used within \name, we can achieve different semantic 
objectives. For example, we may want to construct images that look like 
one category, but are classified as another (as in adversarial examples),
and yet are still plausible. The setting here is similar to %
Song et al. \cite{Song2018ConstructingUA}, but with different generative models and optimization methods.
Using the fitness function
in Equation \ref{eq:bound_fitfunc},

\begin{equation}
  f(i) = 
  \begin{cases}
    \infty & N_c(i) = N_c(i_0) \\
    |E(i)-E(i_0)| & else \\
  \end{cases}
  \label{eq:bound_fitfunc}
\end{equation}

\noindent
we may obtain images that are semantically ambiguous. 
The upper case denotes the case in which the new image's classification has not changed; we
do not want to accept such images. The lower case is when the new image's classification is different
from the original image; in such cases we want the image to look as similar to the original image as possible. 
Specifically, we employ
the following genetic algorithm. A random image $i_0$ is sampled from the test dataset. 
The latent representation of this image, $E(i_0)$, is obtained. The inital population
is constructed by sampling new representations $z_i$, where $z_i = E(i_0) + \epsilon_i$,
and $\epsilon_i \sim N(0, 1)$. 
An image can be reconstructed using the decoder,
so that $i_i = D(z_i)$. Using this image we calculate the fitness of the representation as in Eq. \ref{eq:bound_fitfunc}.
Single-point crossover is used; mutation is performed by adding a small noise vector on each
genome. Individuals with smaller fitness are selected for the next generation. 
An example result of this optimization is shown in
Figure \ref{fig:ambiguous}. Observe that while this image
does look somewhat similar to a
digit, it is not clear whether this image is a 4 or a 9. 
On the
other hand, raw pixel search does not yield semantically plausible results as in Fig. \ref{fig:rawVvae}, 
reducing the utility in terms of analyzing and reasoning about the network. 

\subsection{\rqtwoname (RQ2)}
\label{sec:expr2}
One can suspect that the objective function provided in Eq. \ref{eq:bound_fitfunc} will yield
images close to the decision boundary of corresponding neural networks. To
verify whether this is the case, we utilize dropout~\cite{Srivastava14Dropout}
and measure the variability of predictions. Dropout is a training technique 
that drops a certain proportion of neurons during training; it is known to
make neural networks more robust in general. While dropout is usually used only
when training, we keep dropout on and measure how much predictions vary when
different neurons are dropped. It is likely that images that are close
to the decision boundary will vary in final prediction results
much more than images that are safely within a certain
category's boundary. This bears some resemblance to previous methods that use
dropout to detect adversarial examples~\cite{Feinman2017DetectingAS, Li2017DropoutAdv},
but is simpler than those approaches which incorporate Bayesian optimization.
Using the objective function in Eq. \ref{eq:bound_fitfunc}, we sample 10,000 images
and measure the prediction variability using the following metric:

\begin{equation}
V(I) = \sum_{j=1}^{c} \sqrt{\frac{1}{n-1} \sum_{i=1}^{n} (p_i[j]-\mu_{ij})^2}
\label{eq:pred_var}
\end{equation}

\noindent
where $p_i[j]$ is the $j$th class's predicted confidence in image $i$, and 
$\mu_{ij} = \frac{1}{n} \sum_{i=1}^{n} p_i[j]$. Upon inspection, one can note that
the expression
$\sqrt{\frac{1}{n-1} \sum_{i=1}^{n} (p_i[j]-\mu_{ij})^2}$,
which appears in Equation \ref{eq:pred_var},  
is essentially the
Bessel corrected standard deviation of each category's predicted likelihood,
summed over all categories. We use this metric as it continuously measures
how much a DNN's prediction varies. Table \ref{tab:pred_variation} shows the comparison
results for this experiment.

\begin{table}[t]
\begin{tabular}{ ccc } 
\hline
 Dataset  & GA-generated & Provided \\ \hline
 MNIST    & 0.7456 & 0.1365 \\ 
 CIFAR-10 & 0.7853 & 0.0202 \\ \hline
\end{tabular}
\caption{A comparison of the variance in prediction between provided test images
and GA-generated `boundary' images. The boundary images vary more in their predictions
than the test images, suggesting that they are closer to the decision boundary.}
\label{tab:pred_variation}
\end{table}

As the table shows, the boundary images cause the classification results of a given
neural network to vary much more than when the same neural network is exposed to
the provided test images. These results indicate that slight changes to the composition
of neurons will significantly alter the final prediction in the case of boundary
images. In turn, this indicates that the boundary images are close to the neural
network's decision boundary itself. 

\subsection{\rqthreename (RQ3)}
\label{sec:expr3}
To answer RQ3, we perform differential testing to uncover hidden 
properties of neural networks. Performing differential testing only requires
a minor change from previous experiments: namely, the fitness function would
be changed to Equation \ref{eq:diff_fitfunc}, where
$N$ and $N'$ are distinct neural network classifiers and $i_0$ is an initial
image.

\begin{equation}
  f(i) = 
  \begin{cases}
    \infty & N(i) = N'(i_0) \\
    m|E(i)-E(i_0)| & else \\
  \end{cases}
  \label{eq:diff_fitfunc}
\end{equation}

\begin{figure}[h!]
  \centering
  \includegraphics[width=\columnwidth]{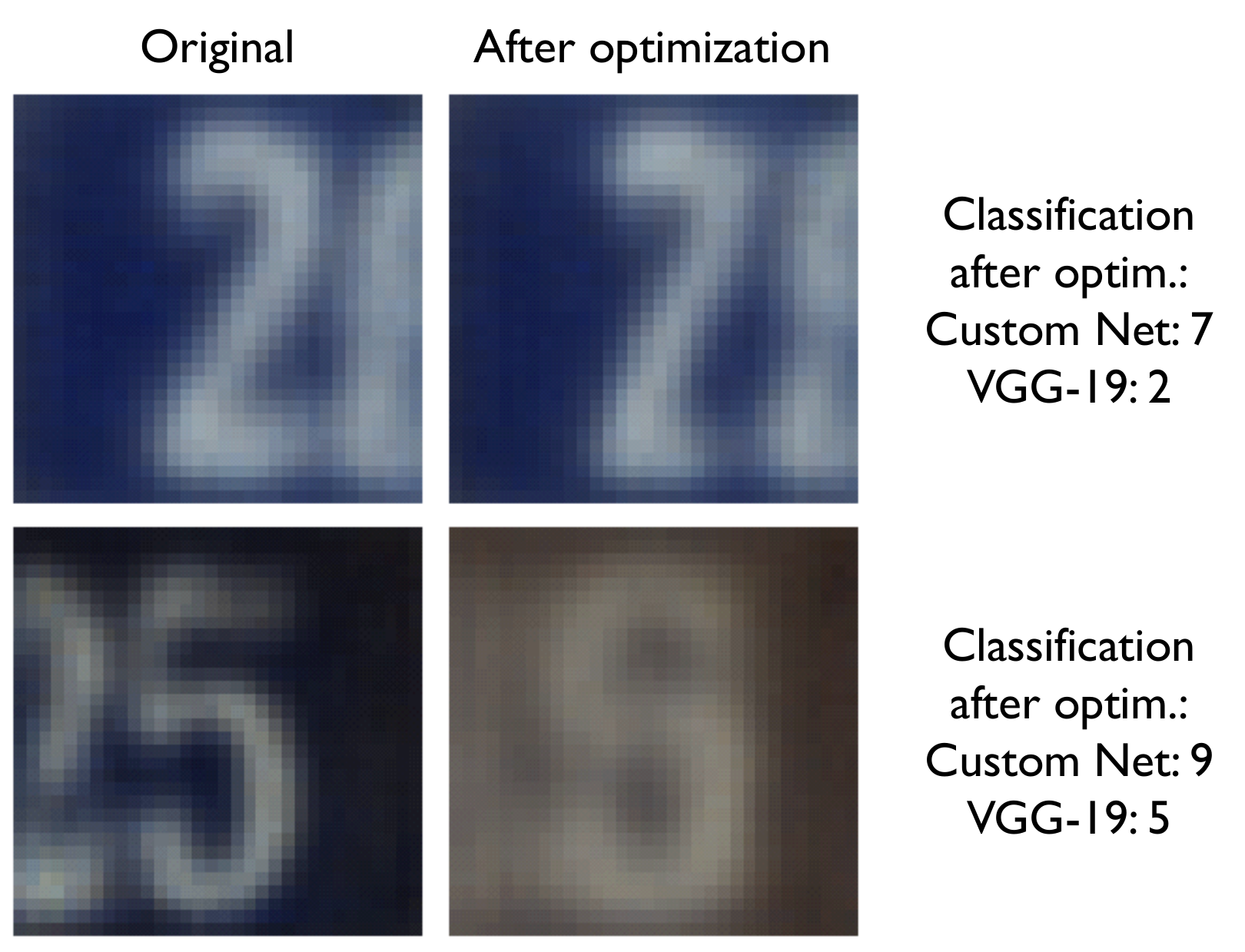}
  \caption{Example images (middle) generated during differential testing of two neural network together with their corresponding classifications (right).
  \name construct images where NNs deviate, while trying to stay more ``realistic''.}
  \label{fig:differential}
\end{figure}

The genetic algorithm used to find such images is the same one as explained
in Section \ref{sec:expr1}, except that the fitness function is defined
as in Equation \ref{eq:diff_fitfunc}. The multiplication factor 
$m=(2+N_p(i)[c]-{N'}_p(i)[c])$ where $c=N'(i_0)$ in the 
fitness function encourages neural networks to diverge as much as possible, where
two is added to keep the multiplication factor positive. 
If we perform this search in the VAE latent space, we are effectively finding
plausible images that cause neural networks to diverge in decision. Hopefully,
it will be possible to identify the underlying reasons behind different decisions 
in neural networks. 
Results of differential testing performed on the SVHN dataset are shown in 
Figure \ref{fig:differential}. Here, our custom network for SVHN and a VGG-19
network are compared, with architectures described in Section \ref{sec:setup}.

Given many such results, one may be able to discern distinct semantic properties
that different neural networks have, and construct an accurate ensemble of DNNs.
For example, if as a result of comparing network A and B, many gray background images
as in the lower right image of Figure \ref{fig:differential} appear, we may infer
that network A is having a hard time with gray background images. 
While these results are far from conclusive, the images show the potential of \name
to compare cases in which neural networks diverge, opening the possibility
of further analysis of the decision boundaries in neural networks. 

\subsection{\rqfourname (RQ4)}
\label{sec:expr4}
\begin{figure}[h!]
  \centering
  \includegraphics[width=\columnwidth]{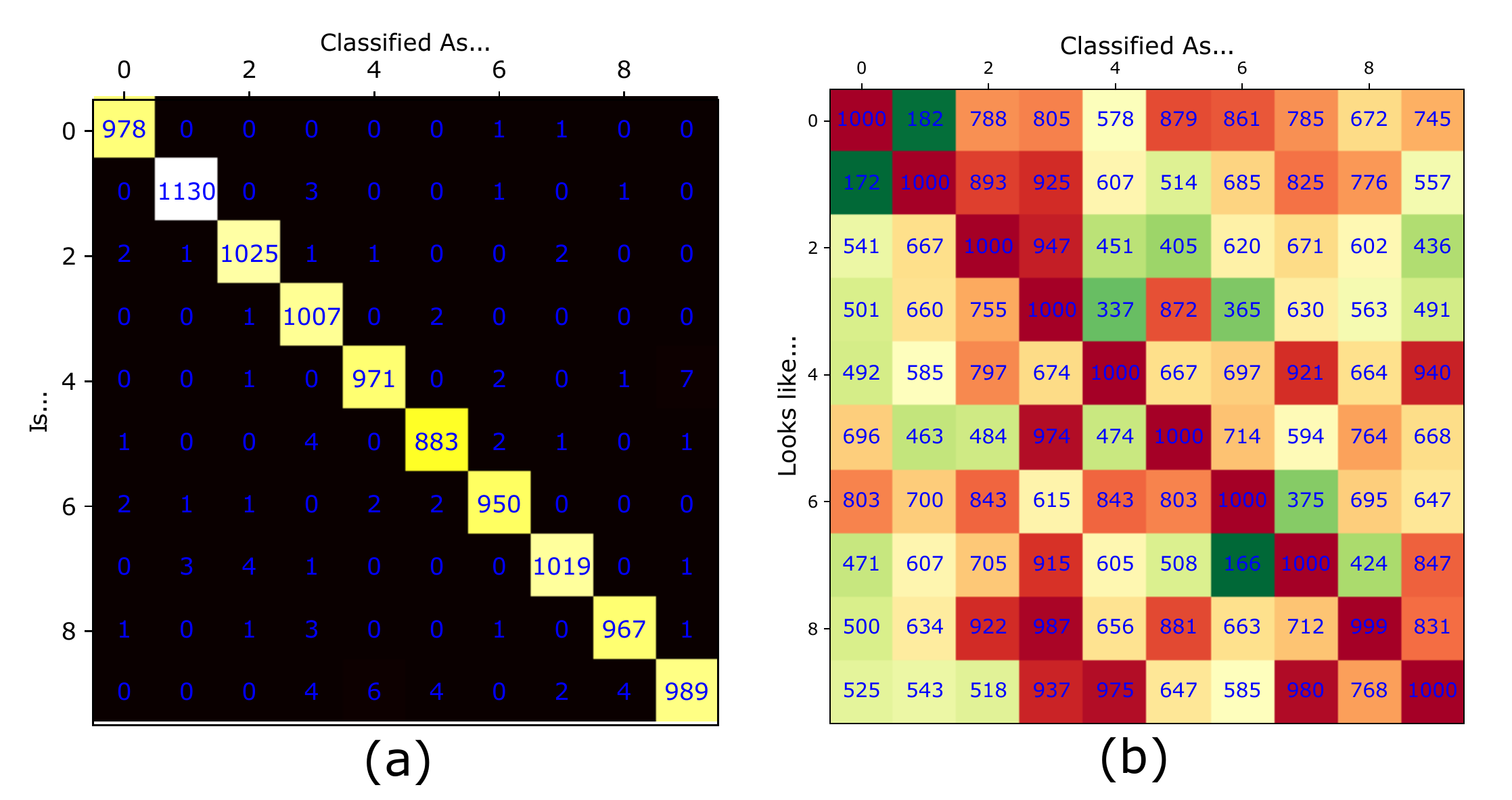}
  \caption{Comparison between (a) the error matrix of the provided MNIST test set
  and (b) the number of successful optimizations of the GA algorithm out of 1000 attempts. Note that the differences are more pronounced in the GA algorithm's matrix. 
  }
  \label{fig:testVfail_mat}
\end{figure}

Using \name, we attempt to find pairs of categories that a given neural network
may be confused by. 
Specifically, a targeted fitness function is devised, and
presented in Equation \ref{eq:target_fitfunc}, where $t$ is a target class. 
\begin{equation}
  f(i) = 
  \begin{cases}
    \infty & \nu(i) \neq t \\
    |E(i)-E(i_0)| & else \\
  \end{cases}
\label{eq:target_fitfunc}
\end{equation}
In fact, this optimization quite often fails, i.e. the search does not find an image 
that has the target class $t$ starting from the image $i_0$.
What is of interest is not that
it fails, but how often, i.e. the proportion of failures. It turns out that image pairs that
are confused between each other have a higher success rate, while image pairs
that are clearly distinguished have a lower success rate. 
Concretely, we define a \textit{GA class escape rate} as 
the number of times an attack
successfully finds a solution that causes the DNN to classify an image as the 
target class divided by the number of all attempts. We can compare this to the
targeted error rate which is the number of times an image in the dataset test set 
with the source class is classified as the target class. 
The change in GA class escape rate is more pronounced than the targeted error rate of the test 
cases, as is showcased in Figure \ref{fig:testVfail_mat}, by juxtaposing the 
confusion matrix using provided test cases and the GA class escape rate for a 
simple neural network classifier. 

\begin{figure}[h!]
  \centering
  \includegraphics[width=\columnwidth]{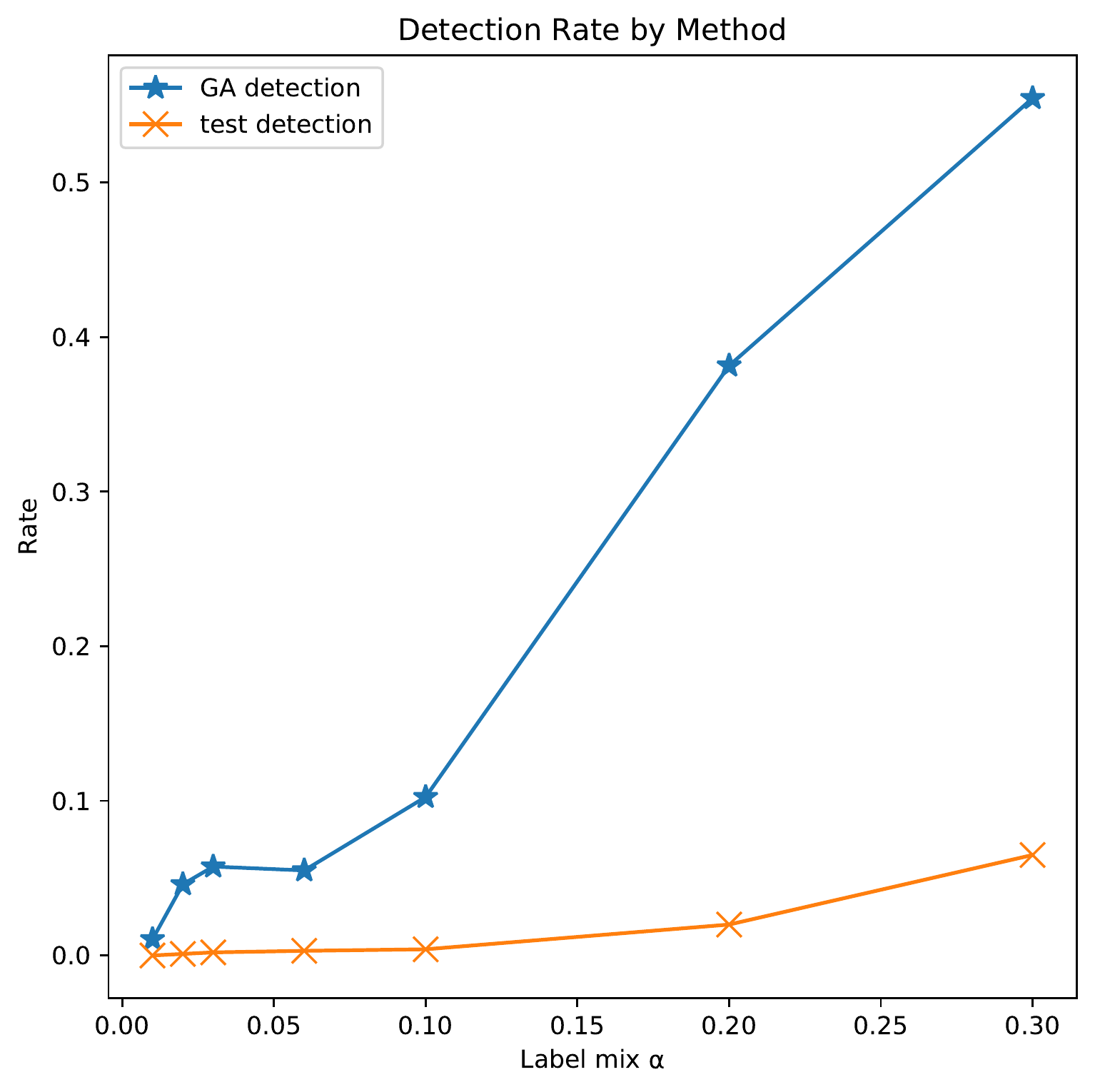}
  \caption{Performance change when labels are mixed. GA class escape rate
  increases more rapidly than error rate when labels are mixed.}
  \label{fig:testVfail_dr}
\end{figure}

To show the effectiveness of the GA class escape rate, target labels are artificially mixed during
the training phase according to a proportion $\alpha$. For each $\alpha$,
a corresponding neural network is trained. For each such neural network, we
measure the error rate between the mixed pair and the GA class escape rate. Results where
0 and 1 are mixed with varying $\alpha$s are presented in Figure \ref{fig:testVfail_dr}.
As evident from the figure, GA class escape rate has a much tighter
correlation relationship with the gradual mixing of labels; default test cases only
pick up this change when labels have been mixed quite a lot. This shows the effectiveness
of \name-generated test cases in finding potential flaws in neural networks.

\section{Threats to Validity}
\label{sec:threats}

\paragraph{Internal Validity.} Internal validity regards whether there are
experimental factors that may influence the conclusions
reached in the paper. 
The results in Section \ref{sec:expr2} and \ref{sec:expr4} are somewhat stochastic 
and values may change in a reproduction experiment. To mitigate this issue, we ran 
these experiments over a large number of images ($n\geq1000$) so that we could be fairly 
confident that the overall trend of these results would be stable despite their 
randomness.

\paragraph{External Validity.} External validity is threatened when the study's
results may not generalize. In this work, we used a small number of neural network
architectures for convenience. As such, it has not been verified whether the results
of this study will generalize to other architectures.
Small and simple image datasets such as MNIST or SVHN are employed instead of larger and more
complex datasets such as CIFAR-100 or the ImageNet dataset; this was a decision
made mostly out of convenience. Additional experiments are required
to verify our results on other datasets.

\section{Discussion and Conclusion}
\label{sec:conclusion}
In this paper, we introduce \name, a technique that uses generative models 
to focus the search for images so that more plausible\slash realistic ones 
can be generated.
Through experiments, we demonstrate that \name can provide potentially more useful
inputs when assessing a neural network, both quantitatively
for various testing scenarios, and qualitatively via visualization and human input.
We also show that, coupled with search, we can find images that
are close to decision boundaries, which can be used for boundary value testing and
for diagnosing whether the boundary is actually where it is expected to be. 
Finally, we show how \name can provide unique insights 
into the behavior of neural networks through differential testing and targeted class 
escape rate analysis.

While the paper focuses on image classifier neural networks, future 
work should also target other data modalities, generative models, and types of machine 
learning models.
The specific generative model used, Variational Autoencoders, have been 
applied to non-image data such as text~\cite{Fanny2018AudioVAE}. 
Moreover, techniques to use them for multi-modal datasets~\cite{sadeghi2019Audiovisual} 
or test suites~\cite{Reichstaller2017VAETestSuite} have been proposed. 
Thus, future work should investigate if generative models coupled with search are a  
general method for generating semantically meaningful software inputs and data.
As other and more powerful generative models are introduced they can also be used.
Similarly, we aim to investigate how our approach can be used to test machine learning models 
other than neural networks. As long as the fitness function we use in our search 
is agnostic to the model under test, i.e. it does not rely on internal 
computations of the model, our method should generalize basically unchanged.

In summary, we believe that the shifting of focus from low-level, ``syntactic'', 
here pixel-level, perturbations to semantically meaningful changes that search coupled 
with \name allows can benefit many interesting software engineering and testing applications. 

\section*{Acknowledgement}

This work was supported by the Engineering Research Center Program through the National Research Foundation of Korea funded by the Korean Government (MSIT) (NRF-2018R1A5A1059921), Institute for Information \& communications Technology Promotion grant funded by the Korean government (MSIT) (No.1711073912), and the Next-Generation Information Computing Development Program through the National Research Foundation of Korea funded by the Korean government (MSIT) (2017M3C4A7068179).

\bibliographystyle{ACM-Reference-Format}
\bibliography{newref}

\end{document}